\documentstyle [12pt]{article}
\def\gsimeq
{\hbox{\raise0.5ex\hbox{$>\lower1.06ex\hbox{$\kern-1.07em{\sim}$}$}}}
\def\lsimeq
{\hbox{\raise0.5ex\hbox{$<\lower1.06ex\hbox{$\kern-1.07em{\sim}$}$}}}
\def\pn{\par\noindent}

\def\ms{\medskip\pn}
\def\bs{\bigskip\pn}

\begin{document}

\renewcommand{\thefootnote}{\fnsymbol{footnote}}

\title{Is the X-ray spectrum of the Seyfert 2 galaxy\\
NGC5252 intrinsically flat ?}

\author{M. Cappi$^{1}$, T. Mihara$^{1}$, M. Matsuoka$^{1}$, W. Brinkmann$^2$,
\\
 M. A. Prieto$^2$, G.G.C. Palumbo$^{3,4}$ }
\date{ }

\maketitle
\pn
   $^1$ The Institute of Physical and Chemical Research (RIKEN),
2-1, Hirosawa, Wako, Saitama 351-01, Japan

\pn
   $^2$ Max-Planck-Institut f\"ur extraterrestrische Physik, D-85748
Garching bei M\"unchen, F.R.G.

\pn
   $^3$ Dipartimento di Astronomia, Universit\`a di Bologna, Via
Zamboni 33, I-40126 Bologna, Italy

\pn
   $^4$ Istituto Te.S.R.E., CNR, Via Gobetti 101, I-40129, Bologna,
Italy

\vspace{1truecm}
Accepted for publication in the Astrophysical Journal (Main Journal).
\vspace{1truecm}
\pn
Send correspondance to:
\pn
Massimo Cappi
\pn
The Institute of Physical and Chemical Research (RIKEN)
\pn
Hirosawa 2-1, Wako-shi,
\pn
Saitama 351-01, JAPAN
\pn
Tel:+81-48-462-1111 ext.3221
\pn
Fax:+81-48-462-4640
\pn

\vfill\eject

\begin{center}
{\large \bf ABSTRACT}
\end{center}
The first X-ray observation of the Seyfert 2 galaxy NGC5252 is reported.
ASCA collected $\sim 4000$ photons/detector enough to perform an
accurate spectral analysis of this source. The luminosity of NGC5252 is
L$_{\rm X}$(0.7-10 keV) $\simeq 2.6 \times 10^{43}$ erg s$^{-1}$,
typical of a Seyfert 1 galaxy.  A simple description of the spectrum
with a single power law is ruled out by the SIS data which shows a
strong soft excess at E $\lsimeq$ 1.2 keV.  The spectrum is best
fitted by models assuming either partial covering of the central source
or scattering of the X-ray continuum.
The best-fit partial covering model results in a flat ($\Gamma \simeq 1.45
\pm 0.2$) power law continuum emitted by a source
almost completely covered (at $\sim$ 94-97\%) by neutral matter
($N_{\rm H} \simeq (4.3 \pm 0.6) \times 10^{22}$ cm$^{-2}$). The
detected iron line is remarkably weaker (EW $\sim 90 \pm 60$ eV) than
normally found from Seyfert 2 galaxies. In an effort to
interpret the observed flat spectrum with an intrinsically steep power law
as predicted by unified models, we also included in the models
neutral reflection, ionized absorption (warm absorber), and
non-uniform cold absorption (dual-absorber).
We find that the observed flat continum {\bf and} weakness of the iron
line poses a problem for neutral reflection models, whatever the
assumed geometry is. The use of an ionized absorber
(scattering + warm-absorber model) instead of a neutral absorber seems
not justified with the present data.
The ``dual-absorber'' model, representing nonuniform (density)
absorption along the line of sight,
may provide an alternative explanation for the
flatness of the spectrum, with no constraints on the Fe K emission line.
Another plausible interpretation of the present data is that the X-ray
spectrum of NGC5252 is truly intrinsically flat. Its strong similarities
with the well-known Seyfert 1.5 NGC4151 are briefly discussed.

\bigskip
{\bf Key-words}: galaxies: individual (NGC5252) - galaxies: Seyfert -
X-rays: galaxies

\vfill\eject

\section {Introduction}

The discovery of broad emission lines in the polarized optical
spectra of several Seyfert 2 galaxies (Antonucci \& Miller 1985,
Miller \& Goodrich 1990, Tran et al. 1992) has provided the basis for
a unified model of Seyfert galaxies in which the main discriminating
parameter between Seyfert 1 and Seyfert 2 nuclei is the inclination to
our line of sight of an obscuring torus surrounding the
central source (see Antonucci 1993 for a review).
In this scheme, Seyfert 1 galaxies are active nuclei observed nearly
perpendicularly to the torus plane (unabsorbed) whereas Seyfert 2 galaxies
represent those seen through the torus (absorbed). Further support of
this scenario has come from GINGA X-ray observations which revealed
the strongly absorbed spectra of some Seyfert 2 galaxies (Koyama et al.
1989, Awaki et al. 1991).
\pn

NGC5252 is an S0 galaxy at a redshift of z $\simeq$ 0.0234.
Its optical size is $\sim
\ 1.4 \ \times$ 0.9 arcmin, with the major axis being at a position angle
PA $\sim$ 15$^{\circ}$, and the absolute visual magnitude is M$_v$ = -21.4
(Unger et
al 1987). NGC5252 harbours a Seyfert 2 nucleus which illuminates a
perfectly defined biconical structure of ionized gas extending up to
35 kpc from the nucleus (Tadhunter \& Tsvetanov 1989). This is one of
the largest extended ionized gas regions among nearby AGN. The axis
of the bicone lies roughly at PA $\simeq$ 166$^{\circ}$, close to the
main axis of the 20 cm radio structure ($\simeq$ 172$^{\circ}$).
VLA HI observations of this galaxy revealed the existence of neutral
gas located just outside the bicone of ionized gas. These findings
provided unambiguous proof of the anisotropy of the radiation
field in this galaxy (Prieto \& Freudling 1993).
Additional radio
observations have detected weak, narrow extended features aligned
with the larger scale optical ionization cones which further support
the existence of collimation effects in NGC5252 (Wilson \& Tsvetanov 1994).

 Because of its favourable orientation with respect to the observer,
NGC5252 is an ideal target for studying the collimation mechanism
responsible for the observed anisotropy. Combined near IR and optical
imaging have revealed the presence of a band of material across the
nucleus and perpendicular to the bi-cone of ionized gas (Kotilanien \&
Prieto 1994). The band shows a gradually increasing extinction
towards the center of the galaxy; a lower limit to the extinction of 6
mag is estimated.

Near IR spectroscopy of HeI 1.083 $\mu$m by Ruiz, Rieke \& Schmidt
(1994) shows indeed a broad HeI component with FWHM $\sim$ 1050 km
s$^{-1}$ thus revealing the typical signature of a Seyfert 1 nucleus.
Polarization studies aimed at discovering the nature of the obscured
nucleus via the nuclear light scattered by the surrounding material
are not yet conclusive. Optical spectropolarimetry of the galactic
nucleus indicates a (11 $\pm$ 11)\% polarization (Kay 1994). Given the
strong (larger than 90\%) stellar contribution to the nuclear optical
light in this Seyfert 2 galaxy, polarization dilution by stellar light
could be significant (Kay 1994; Kotilanien \& Prieto 1994);
alternatively, scattered light could also suffer extinction.

At the opposite extreme of the electromagnetic spectrum, broad-band
X-ray data can provide crucial information on the hidden source because
photons of a few keV may pass through the nuclear intervening
material. It should be noted that since NGC5252 is an early-type galaxy,
the contribution from star formation activity to the X-ray emission is
expected to be minor. Indeed, deep H$\alpha$ imaging (Prieto \& Freudling,
in preparation) shows no trace of ionized gas
outside the [OIII]$_{5007 \AA}$ ionized bi-cone reported by Tadhunter
\& Tsvetanov (1989); most of this gas extends out of the
galaxy plane.  NGC5252 was not detected in the ROSAT all-sky
survey.  Based on these general considerations, we investigated the
X-ray properties of NGC5252 with the ASCA observatory.
To our knowledge, there is no report
of any X-ray detection from previous missions. During the ASCA
observations, enough photons were collected (section 2) to allow a
detailed and accurate spectral analysis (section 3). This analysis is
discussed in the light of recent observations of Seyfert galaxies
and theoretical models (section 4). Our main results are summarized in
section 5.

\section{ASCA data reduction}

ASCA observed NGC5252 between January 28 and January 31, 1994 with the
Gas Imaging Spectrometer (GIS) for a total effective exposure time of
about 70500 s and with the Solid-state Imaging Spectrometer (SIS) for
about 65000 s. The SIS was operating in 1 or 2 CCD mode and all the
data were collected in FAINT mode. DFE and echo corrections have been applied
to the data (Otani \& Dotani 1994). Good observational intervals were
selected by applying the data selection criteria shown in Table 1.

Source counts were collected from a circle centered on the source
of 4 arcmin radius for the GIS and 2.5 arcmin for the SIS. Background
spectra were obtained using similar areas inside the Field of View (FOV) from
uncontaminated regions. We checked for differences between different extraction
regions for the source and the background but did not find any
significant variations in the spectral results. In the GIS FOV, we
detected elongated spurious emission in the direction north-west of NGC5252.
This emission occurs in the direction of the ionized cones but
we can exclude the possibility that it is related to NGC5252 because of the
high
separation ($\sim 10$ arcmin) between the two. It is more likely
associated with the bright SB galaxy MCG-01-35-024 whose position is
coincident with the emitting region to within $\sim$ 2 arcmin.
We emphasize that the use of blank sky backgrounds for the GIS and SIS yielded
spectral results within the errors reported in the following spectral
analysis.

Relevant data for the NGC5252 observation are summarized in Table 2.
Data preparation and analysis has been done using version 1.0h of the
XSELECT package; spectral analysis has been performed using version
8.5 of the XSPEC program (Arnaud et al. 1991).

\section{Spectral analysis}

Source plus background light curves were computed in different energy
bands: in the broad (0.4-10 keV), the soft (0.4-2 keV) and the hard
(2-10 keV) energy bands.  None of these lightcurves indicated
significant variability; statistical $\chi^{2}$-tests against
constancy give a probability $>$ 15\% in every case, i.e.,
consistent with constant source flux.

Therefore, we accumulated spectra from the photons collected in the total
$\sim$ 70 ks
observation for the spectral analysis.  The pulse height
spectra were binned in such a way that there were at least 30 and 20
counts/bin for the GIS and SIS detectors and the corresponding energy
bands were limited to 0.7-10 keV and 0.4-10 keV, respectively.  We used
the detector response matrices gisv3\_1.rmf (GIS, released in March
1994) and rsp1.1.alpha (SIS, released in June 1994) and constructed
the effective area at the source position with the program
jbldarf-0.5a (released in January 1994).

\subsection{Single power law fits}

A single absorbed power law model was first fitted to the data with the
effective hydrogen column density ($N_{\rm H}$) left free to vary. The
resulting
best-fit parameters are given in Table 3. This simple model yielded an
acceptable description of the GIS data in the 0.7-10 keV band with
$\chi^{2}_{red}/{\rm d.o.f.}\simeq 1.03/267$ but gives an unacceptable
fit to the SIS data in the 0.4-10 keV band, with $\chi^{2}_{red}/{\rm
d.o.f.} \simeq 1.38/315$. The corresponding SIS spectrum is shown in
Figure 1.  The residuals of the SIS spectrum clearly indicate the
presence of strong soft (E$\lsimeq$ 1.2 keV) X-ray emission in
excess to the extrapolation of the higher energy continuum.
It should be pointed out that the discrepancies between the GIS and
SIS results (Table 3) are strongly reduced when the soft excess is
explicitly fitted in the SIS spectrum as shown in the following
sections.  In all fits, the best-fit value of $N_{\rm H}$ is at least
2 orders of magnitude higher than the galactic absorption $N_{\rm
Hgal}$ $\simeq$ 1.97 $\times 10^{20} $ cm$^{-2}$ (Dickey \& Lockman
1990), suggesting strong intrinsic absorption.

It should be emphasized that, for completeness, an absorption
component with $N_{\rm H}$ = $N_{\rm Hgal}$ was also added to each
model tested below in order to account for the absorption in our own
galaxy. Because of its low value, however, the Galactic column is effective
only at energies E $\lsimeq$ 0.4 keV.

\subsection{Soft excess}

\subsubsection{Thermal soft-emission models}

In a first attempt to model the soft X-ray excess, the broad-band pulse height
spectrum was fitted with the following two-component models: (a) black
body + absorption $\times$ power law, (b) bremsstrahlung + absorption
$\times$ power law and (c) Raymond Smith + absorption $\times$ power law,
with the absorption applied to the power law free to vary in all models.

In model (a), the black body emission component could
mimic the high energy tail of the spectrum emitted by an accretion
disk.  In models (b) and (c), it is assumed that the soft X-ray
emission is produced by a thermal thin and hot gas associated with the
host galaxy.  In all three models, the hard power law is thought to be
related to the nucleus itself.  With all parameters free to vary,
these models yielded acceptable fits to the SIS data with
$\chi_{red}^{2} /$d.o.f. $\sim$ 0.93-1.1/311.  Our results are
reported in Table 4. All these models yielded similar values of
$N_{\rm H}$.

Fitting with model (a) gave L$_{\rm X}$(0.1-2.0 keV)$\simeq$
3.9 $\times$ 10$^{41}$ erg s$^{-1}$ and a black body temperature
of $T \simeq$ 2 $\times$ 10$^{6}$ K. From these values, we deduce a
characteristic dimension of the emitting region of r $\simeq 3\times 10^{9}$
cm which is 2 orders of magnitude smaller than the Schwarzschild
radius for a $10^{6}\ M_{\odot}$ black hole. Therefore, a black-body +
power law model does not provide a physically meaningfull description of the
data.

Thermal bremsstrahlung (model b) with a characteristic temperature $\sim$ 0.8
keV together with a hard power law with $\Gamma \simeq 1.3 \pm 0.2$ provides a
possible description of the data.
However, this interpretation is inconsistent with the lack of soft emission
lines in the ASCA spectrum which are expected to be emitted by a
hot ($T \sim 10^7$ K) plasma. The same problem is faced by a
Raymond-Smith model (model c; Raymond \& Smith 1977).
In this case, metal abundances below $\sim$ 0.01 solar are needed for
acceptable fits, as reported in Table 4.
These results, together with general arguments concerning the host
galaxy (see section 4.1), make the hypothesis of a thermal origin for the
soft-excess unlikely in the case of NGC5252.

\subsubsection{Partial covering model}

The observed SIS spectrum was then fitted with a partial covering
model (Holt et al. 1980). This model consists of a power law
continuum obscured by cold uniform matter with column density $N_{\rm H}$,
which covers a fraction
$C_{\rm F}$ of the source.  The model has 4 free
parameters, namely the normalization, the photon index, the covering
fraction and the column density.  In physical terms, the absorbing
material might be identified with thick and cold clouds distributed
along the line of sight, covering only part of the central source.

As illustrated by the unfolded spectrum in Figure 2, this model
provides a good description of the SIS (0.4-10 keV) spectrum with
$\chi^{2}_{red}/{\rm d.o.f.} \simeq 0.95/315$.  The best-fit
parameters are reported in Table 5a. Confidence contours for the photon
index versus column density and for the covering fraction versus column
density are shown in Figure 3.
The best-fit parameters show that the SIS spectrum of
NGC5252 is consistent with a flat ($\Gamma \simeq 1.45 \pm 0.2$)
underlying power-law viewed through an absorber with column
density $N_{\rm H} \simeq (4.3 \pm 0.6)\times 10^{22}$ cm$^{-2}$
which covers $\sim 96$ \% of the primary source. The absorption
corrected flux between 0.4-10 keV is 11.7 $\times$ 10$^{-12}$ erg
cm$^{-2}$ s$^{-1}$.  We emphasize that the reported errors correspond
to 90\% confidence limits for 3 interesting parameters and are,
therefore, very conservative limits (Lampton et al. 1976).
\pn
The derived $N_{\rm H}$ value is equivalent to Av $\sim$ 20 mag, for a
standard gas to dust ratio, and thus is consistent with the large
extinction towards the central source predicted from the IR-optical data
(Kotilanien \& Prieto 1994).

It should be pointed out that with the partial covering model the shape
of the power-law in the SIS is consistent with the results obtained from the
GIS instrument with a single power law model (section 3.1).  Fitting
the GIS data with a partial covering model, one obtains best-fit
parameters which agree with the SIS results within $\sim$ 3\%.

\subsubsection{Scattering model}

Another model which could describe the soft excess is a
scattering model. Numerically equivalent to the partial covering model,
it consists of the sum of 2 power laws having the same photon index but
different absorptions and normalizations.  One power law is absorbed by the
Galactic column density only and
the other is absorbed by a high intrinsic column density (a free
parameter during the fit procedure).  In this model the
``non-absorbed'' and absorbed power laws are related to the presence
of scattering and obscuring matter, respectively.  This interpretation
is supported by unified models of AGN which predict
that the direct continuum of Seyfert 2 galaxies is strongly obscured by a
thick torus (Ghisellini, Haardt \& Matt 1994) whereas part of the nuclear
emission
is scattered by free electrons surrounding the BLR (Antonucci \&
Miller 1985, Awaki et al. 1991).

As reported in Table 5b, the scattering model yields a good
description of the data with $\chi^{2}_{red}/{\rm d.o.f.} \simeq
0.95/312$, the best-fit parameters being again a flat power law with $\Gamma
\simeq 1.46 \pm 0.2$, intrinsic absorption $N_{\rm H}
\simeq (4.4 \pm 0.6) \times 10^{22}$ cm$^{-2}$ ($\sim N_{\rm H}$ partial
covering),
and a fraction of $\simeq$ 4-5\% of the continuum scattered into the observer's
line of sight.

\subsection{Flat spectrum}

On one side, $HEAO-1$ A2 (Weaver, Arnaud \& Mushotsky 1995) and $Ginga$ (Nandra
\& Pounds
1994) results have shown that the {\it observed} canonical
photon index $\Gamma_{2-18keV} \sim 1.7 - 1.8$ of Seyfert 1 galaxies can be
interpreted as the sum of a steep {\it intrinsic}
power law with $\Gamma_{2-18keV} \sim 1.8 - 2.1$ and a reflection
component (e.g., Lightman \& White 1988).
On the other side, we still don't know very much about Seyfert 2 galaxies,
as they are generally more absorbed and only few of them have high
S/N X-ray spectra.
Seyfert unification models (e.g., Antonucci 1993) predict the same
average {\it intrinsic} photon index for Seyfert 1 and Seyfert 2 galaxies, but
there is increasing evidence that the {\it observed} photon indices of
several Seyfert 2 galaxies are flat with
$\Gamma_{2-10keV} \sim 1.4 \pm 0.2$ (Awaki et al. 1991, Kunieda et al. 1994).
Whether the {\it intrinsic} photon index of these is flat or not
is an important question that directly impacts Seyfert unification models.

We have shown that all the models applied to NGC5252 require a flat power law.
In light of the above considerations, it is therefore tempting to use more
complex models
to search for a global description of the spectrum with an intrinsically steep
power law,
as predicted by unified models.
Therefore, in the following sections, we will focus on models involving (a)
reflection
by thick material, (b) absorption by ionized matter, and (c) line of sight
obscuration.
These models will be used together with the partial covering model (a and c) or
scattering model (b). It should be borne in mind that though they imply
different physical interpretations, the partial covering and scattering model
cannot
be distinguished with the present data (section 4.1) and would, therefore, give
equivalent results if interchanged.

\subsubsection{Reflection component and composite models}

We first considered the possibility of explaining the apparently flat
spectrum of NGC5252 with the presence of a reflection component. The
existence of cold gas responsible for radiation reprocessing
near the center of AGN is suggested by the strong UV
emission in many observed AGN and is supported by theoretical considerations
(Guilbert \& Rees 1988).  It is well established that the presence of
this reflecting matter can strongly modify
the observed X-ray spectrum. In particular, it could flatten the
observed continuum (Lightman \& White 1988) and
provide an excellent explanation for at least part, if not all, of the
iron emission line intensity.

The reflection model used in the present analysis is the general case
of neutral reflection from a slab of optically thick matter with $\tau_{T} \gg
1$ proposed by Lightman and White (1988), with opacities taken from Morrison
and McCammon (1983). The spectrum is calculated integrating over all
viewing angles and assuming isotropic emission from the source.
The model used is the ``plrefl'' reflection model available
in the XSPEC fitting package (Arnaud et al. 1991).
This model requires a total of 5 parameters, namely the
photon index ($\Gamma_{int}$) and normalization ($A_{\rm pl}$) of
the primary power law, the normalization of the reflected
component ($A_{\rm refl}$), the slab inclination ($i$) and
the fraction of the solid angle subtended by the slab ($\Omega /
2\pi$). The latter parameter ($\Omega /2\pi$) has been fixed
to 1 because it is redundant with $A_{\rm refl}$ in the fitting procedure.
The inclination $i$ was fixed to 0 (slab assumed face-on) because
the shape of the reflected continuum is only a weak function
of the inclination. In its simplest form, the one used here,
this model is completely determined by only 3 (free)
parameters ($\Gamma_{int}$, $A_{\rm pl}$ and $A_{\rm refl}$).
By examining the ratio $R$ $\equiv A_{\rm refl}$/$A_{\rm pl}$, we
can then make inferences about these parameters. For example, a value of $R
\sim$ 1
is consistent with an isotropic source covered
by $\Omega /2\pi \sim$ 1. A value of $R >$ 2
cannot be explained with only an increase in $\Omega/2\pi$, but requires
a time lag effect or/and anisotropic emission from the source.

This model was combined together with the partial covering model
(section 3.2.2), yielding a total of 5 free parameters (one
additional one if compared to the single partial covering model).
The photon index was left free to vary. The resulting best-fit
parameters are reported in Table 6 and the best-fit model is
shown in Figure 4. The inclusion of a reflection component
improves the fit by $\Delta \chi^2 \sim$ 6, which is
significant at more than 95\% significance level for the addition of one free
parameter.
The resulting amount of reflection
is $R \simeq 7.2^{+7.6}_{-5}$ and the photon index is $\Gamma
\simeq 1.91 \pm 0.33$.
The errors quoted here and below in this work
represent 90\% confidence intervals for 1 interesting parameter
($\Delta \chi^{2}$ = 2.701, Avni 1976).
It is clear from these values and from Figure 4 that a very strong
reflection component is necessary if one wants to model the observed
flat shape at energies 4 $<$ E $<$ 10 keV with an underlying
steep power law.

The data are therefore consistent with an intrinsically
steep ($\Gamma \sim 1.9$) power law provided there is a very strong
($R \sim 7$) reflection component. However, as will be discussed later
(section 4.2), the presence of such a strong reflection
is unlikely, though not ruled out, in NGC5252 because it predicts
a stronger than observed iron K$\alpha$ line (section 3.4).

\subsubsection{Warm absorber and composite models}

We further investigated the possibility that a steep ($\Gamma = 1.9$) power
law passing through a warm absorber can give an apparently flat
0.4-10 keV spectrum.
In contrast to a neutral absorber (where we use a partial covering model),
this model interprets absorption as being due to highly ionized gas that
completely covers the continuum source.
The presence of absorption due to ionized matter has currently been confirmed
in several Seyfert 1 galaxies (see Matsuoka 1994 for a review) and may also
contribute significantly to the soft X-ray emission of intermediate
Seyfert galaxies such as NGC4151 (Weaver et al. 1994b).
We therefore investigated whether a warm absorber can explain both the soft
excess and flat spectrum of NGC5252.

A simple model of a warm absorber using the ``CLOUDY'' photoionization code
(Ferland 1991) has been constructed. Following Fabian et
al. (1994), we assumed that a shell of gas with solar abundances,
density $10^{8}-10^{10}$ cm$^{-3}$, and uniform temperature is uniformly
distributed around an ionizing power law continuum source with $\Gamma$
= 1.9 and luminosity L $= 10^{43}$ erg s$^{-1}$.  The emergent
spectrum, consisting of the transmitted continuum plus the emission
from the shell, was fitted to the SIS data.  The free parameters
were the ``warm'' column density $N_{\rm W}$ (cm$^{-2}$), the
ionization parameter $\xi = {\rm L}/nR{^2}$ (erg s$^{-1}$ cm) and the
normalization constant of the warm absorber model.
The model gave a poor fit to the data (with
$\chi^{2}_{red}/d.o.f.=1.32/314$) and strong systematic differences
between the model and the data were evident between 0.7 and 1.6 keV.
Indeed, the warm absorber hypothesis does not work unless another soft
component
is added because the strong edges around 1 keV predicted by the
model (Netzer 1993, Krolik \& Kriss 1995) are not seen in the data.

We then repeated the fitting procedure with an additional power
law to model a soft scattered component (increasing the
number of free parameters to 5).
The photon index of the warm absorbed power law was
again fixed to 1.9 but that of the scattered component was free to vary.
All other parameters were also varying.  The resulting composite
scattering + warm absorber model yielded an acceptable fit with
$\chi^{2}_{red}/d.o.f. \simeq 0.96/311$. The best-fit model is shown
in Figure 5 and the interesting best-fit parameters are log $N_{\rm W}
\simeq 22.88 \pm 0.02$, log $\xi \simeq 1.61 \pm 0.04$ and $\Gamma
\simeq 1.43^{+0.41}_{-0.73}$ for the scattered component.
For two interesting parameters, the 90\% upper limit for the photon index
of the scattered component is $\sim$ 1.92.

Thus, when the spectrum of NGC5252 is modelled with a composite
warm absorber plus (scattered) power law, an intrinsic index
as steep as 1.92 cannot be ruled out.
However, despite the higher number of free parameters, the
statistics are worse than with the single partial covering
model, probably because of the presence of systematic positive
residuals at high ($E > 6$ keV) energies (Figure 5).


\subsubsection{Composite ``dual-absorber'' model}

A further possibility to mask a steep intrinsic continuum slope
is through the presence of an additional, strongly
absorbed power law.  This could considerably flatten the observed spectrum in
the ASCA energy band, similar to a very strong reflection
component. We then fitted the data by adding a second absorbed
power law to the partial covering model.  The resulting spectral model
can be interpreted as a single power law undergoing absorption by
material with 2 different column densities that partly, but not
completely, cover the source.  For example, this may be a crude
approximation of a nonuniform density distribution of clouds along the
line of sight.
\pn
The resulting composite ``dual-absorber'' model is of the form:
partial covering absorption $\times$ power law (1) + partial
covering absorption $\times$ power law (2).  Photon indices and
covering fractions were fixed to be equal for both components, leading to a
total of 6 free parameters, namely the 2 different absorptions
($N_{\rm H1}$ and $N_{\rm H2}$), the 2 normalizations A(1) and A(2),
the covering fraction $C_{F}$, and the photon index.

The model yielded an acceptable description of the source spectrum
with $\chi^{2}_{red} /$ d.o.f. $\simeq$ 0.92/311.  The resulting
best-fit parameters reported in Table 7 are consistent with an
intrinsically steep ($\simeq 1.86 \pm 0.3$) power law source almost completely
($\sim 98$\%) covered by two distinct absorption columns with $\sim 4.8
\times 10^{22}$ cm$^{-2}$ (for $\sim$ 55\% of the covering
matter) and $\sim 6.2 \times 10^{23}$ cm$^{-2}$ (for $\sim$ 45\%
of the covering matter).  As is evident from the best-fit model shown in
Figure 6a, the strongly absorbed power law could well
flatten the hard part of the spectrum in the ASCA energy band.
Therefore, this model provides a viable description of the spectrum
of NGC5252 consistent with a steep power law. It will be considered
again in section 4.2.

The results obtained with the scattering model are also shown for
comparison in Figure 6b. As stated above,
the fits are algebraically equivalent, but in the scattering case the
spectral model could be related to a physical scenario where
the continuum is (completely) covered by a torus of nonuniform density,
whereas part of the continuum emission is scattered into the line of sight
by highly ionized matter present in the vicinity of the nucleus.


\subsection{Iron emission line}

We inspected the SIS spectra for the presence of an iron
emission line feature and our results are summarized in
Table 8.


With the continuum described by a thermal component plus an absorbed
power law ($\Gamma \sim 1.2$), as given in section 3.2.1, we obtain an
upper limit for the equivalent width of an iron K$\alpha$ line
of EW$\lsimeq$ 120 eV.
With the single partial covering model ($\Gamma \sim 1.45$, section 3.2.2),
though the feature is only marginally visible in
the data (Figure 2), a narrow emission line is detected. The
inclusion of a narrow line at the fixed energy E=6.254 keV improves
the fit to more than 99.5 \% confidence level ($\Delta
\chi^{2} \simeq 8$ for 311 d.o.f.). With the energy of the line free to
vary, the best-fit value found is EW $\simeq 90^{+63}_{-49}$ eV
when the width is fixed to 0, and EW $\simeq 100^{+62}_{-59}$ eV when the
width is free, with $\sigma \lsimeq$ 118 eV (Table 8).
These results agree with the emission of a fluorescent narrow Fe
K$\alpha$ line from matter in a neutral state (Makishima 1986).

With the continuum modelled by a partial covering plus reflection
component, the line becomes slightly weaker, with EW $\simeq 78^{+44}_{-53}$
eV.
This value is of particular interest because it can be
directly compared with the expected EW from the
reflection component, as shown in section 4.2.

The line is remarkably weak when compared to reported values for
other Seyfert 2 galaxies (e.g., Kunieda et al. 1994).
As far as results from the iron line diagnostics are concerned, there
is no need for ``complex'' geometries in the case of NGC5252. A simple
line-of-sight absorber, uniformly distributed around the source, can
explain the observed iron line intensity of NGC5252 (Makishima
1986, Leahy \& Creighton 1993).
We emphasize that no absorption edge, expected from neutral matter,
has been detected at the redshifted energy of E = 6.94 keV. The
upper limit derived is $\tau \lsimeq 0.16$ at a 90\% confidence level.
At higher energies there is no evidence for a Fe K edge produced by
ionized matter, either.

 \section{Discussion}

 \subsection {Soft excess}

We have shown that the observed strong soft excess in NGC5252 can well be
described by either a scattering model or a partial covering model.
Thermal models are unlikely.

Recent ASCA observations clearly detected thermal emission
associated with starbust activity from the stereotype Seyfert 2 galaxy
NGC1068 (Ueno et al. 1994). A similar component is unlikely in the case
of NGC5252; its host galaxy is an early-type galaxy
which doesn't exhibit strong stellar activity. More specifically, given
a far-infrared luminosity of $\sim$ 6 $\times$ 10$^{9}$ L$_{\odot}$, the
fitted soft X-ray luminosity of Lx(0.4-2 keV) $\sim$ 6 $\times$
10$^{41}$erg cm$^{-2}$ s$^{-1}$ exceeds the value expected from
the far-infrared/X-ray luminosity correlation for bright infrared
normal and starbust galaxies (David, Jones \& Forman 1992) by
approximately 2 orders of magnitude.
This strongly suggests that
the AGN inside NGC5252 is directly responsible for the observed
soft excess and excludes strong stellar activity (e.g., X-ray
binaries and Supernovae Remnants) associated with the host galaxy.
Moreover, a hot thin collisionally ionized plasma at a
temperature of $T \sim$ 9 $\times$ 10$^6$ K should produce a
blend of emission lines (mostly from Fe L) in the energy range 0.8 to 1.2 keV,
which
are not detected (section 3.2.1). A one-temperature thermal plasma is
therefore ruled out by the data, unless the metal abundances are very small
($\lsimeq$ 0.01 solar).

As long as the scattering zone is not highly ionized
($T \sim 10^{5}$ K), the scattering model predicts the presence of oxygen
absorption edges between 530 and 870 eV, depending on the ionization
state. An estimate of the expected
oxygen optical depth $\tau_{o}$ due to the scattering material is given
by $\tau_{o} \sim A_{o} f_{scatt} \sigma_{o} (\sigma_{e}
c_{f} \times \Omega/4\pi)^{-1}$ (Weaver et al. 1994a), where
$A_{o}, f_{scatt}, \sigma_{o}, \sigma_{e}, c_{f}$ and $\Omega/4\pi$ are
the oxygen abundance, the fraction of scattered flux, the oxygen and electron
scattering cross sections, the covering fraction, and the solid angle
of the scattering
matter, respectively. From the measured opening angle
of $\sim$ 75$^{\circ}$ for the ionization cones and the best-fit value
$ f_{scatt} \sim 0.04$, assuming $c_{f} = 1$ for the scattering medium
and cosmic abundances, we estimate a value of $\tau_{o} \sim 11$.
However, there is no evidence for such a strong edge in our data. As an
example, the upper limit we measure for the OVIII edge at 0.85 keV
(the strongest at soft energies) is $\tau \lsimeq 1.5$.  These results
make the presence of a $T \sim 10^{5}$ K scattering
matter in NGC5252 unlikely.

If the scattering region is highly ionized, X-rays are
efficiently scattered and no absorption features are expected.
Soft ($\sim$ 900 eV) emission lines, typically with EW $\lsimeq$ 100 eV,
are instead expected for a highly
photoionized gas (Netzer 1993).
In the case of NGC5252, the data do not allow significant detection
of any soft X-ray feature, but cannot rule out soft emission lines
of up to several hundreds eV, either.
The existence of a hot ($T \sim$ few $\times 10^6$ K) gas has been
postulated by Marshall et al. (1993) to explain the higher ionization states
indicated by multiple Fe K and Fe L emission lines observed in the BBXRT
spectrum of NGC1068.
Also in the case of NGC5252, gas inside its
bicone might be highly photoionized since it directly sees the
nuclear source and, therefore, could scatter the X-ray continuum into
our line of sight producing the observed soft excess. Polarization of
the optical-UV continuum light should then be detected along the
ionizing bicone. The recent blue (3200-6200 \AA) spectropolarimetric
observations conducted by Kay (1994) are, however, not very
encouraging as no clear polarization was detected (P $\simeq 11\% \pm
11\%$). As proposed by the author, imaging polarimetry or
spectropolarimetry with the slit aligned along the axis of the cone
could be more fruitful.

On the other hand, the partial covering model (Holt et al. 1980)
doesn't predict any soft X-ray line emission as the observed soft
radiation escapes directly from the source.
Since no X-ray
variability was detected with ASCA, it is impossible to distinguish
between a partial covering and scattering model with the present data.
High spatial resolution X-ray
observations with the HRI onboard ROSAT could easily resolve some
extended emission (if present), or/and
optical-UV polarimetric observations could help to discriminate
between these two models.

\subsection {The flat spectrum problem}

The flat spectrum {\bf and} weak iron line differentiates NGC5252 from
the majority of known Seyfert 2 galaxies.
X-ray spectra of most Seyfert 1 galaxies are found to be consistent
with a steep intrinsic spectrum ($\Gamma_{2-18 keV} \sim 1.8 - 2.1$, when
reflection is accounted for) and relatively strong (EW$\sim 120-160$ eV)
fluorescent iron lines (e.g., Nandra \& Pounds 1994).
GINGA observations of Seyfert 2 galaxies
(Awaki et al. 1991) and recent ASCA observations (e.g., NGC6552: Fukazawa
et al. 1994; NGC1068: Ueno et al. 1994) indicate that several Seyfert 2
galaxies have a flat hard X-ray spectrum and a strong (EW $\sim$
several hundred eV) fluorescence iron line (Kunieda et al. 1994).

These features (flat spectrum and strong iron line), first detected by
GINGA and now confirmed by ASCA, are often interpreted as the
signature of reflection of radiation from cold matter (e.g., Reynolds
et al. 1994). In the case of NGC5252, results from section 3.3.1 show
that the data require a very strong
($R \sim$ 7) reflection component to explain the observed flat spectrum.
This high amount of reflection, if true, implies an anisotropic radiation
field, i.e. the reflector ``sees'' more radiation
than is observed directly (Ghisellini et al. 1991), or/and that there is
a time lag effect. However, we believe that a
substantial contribution from a reflection component is unlikely
because it predicts a stronger iron line than observed.
In fact, assuming cosmic abundances, the equivalent width (EW$_{refl}$)
of the iron line relative to the reflection component (measured at
the line energy) is of the order of $\sim$ 1 keV.
We obtained this result from performing Monte Carlo simulations in
the simplest case of a slab geometry (Lightmann \& White 1988), but
as a rough estimate it holds for almost all geometries and for a wide
range of column densities (10$^{22}$ to 10$^{24}$ cm$^{-2}$) of the
reflecting matter (see Makishima 1986 and Awaki et al. 1991
for calculations with spherical and toroidal
geometries, respectively). With the relationship EW$_{refl} \sim$ 1 keV,
the best-fit reflection model predicts an equivalent width (calculated
with respect to the power law plus reflection continuum) of
$\simeq 400 \pm 250$ eV for the iron line.
This value is about 5 times larger than the value (EW $\simeq 78 \pm$ 50,
section 3.4) measured from the data. One should point out that the two
values are still consistent
within 90\% confidence (for two interesting parameters) because of the
large errors associated with the parameters of the reflection component,
but they exclude each other at the 1$\sigma$ level.
The lack of a strong iron line in NGC5252 suggests that a strong
contribution from a reflection component, as possible explanation of the
flat spectrum, is unlikely.

One possible objection to the above comes from the edge-on geometry of
NGC5252, as inferred from optical observations discussed in section 1.
Our previous assumption of a slab observed face-on (the inclination was fixed
to 0 in
section 3.3.1) may not be correct and a larger inclination
angle might have been more natural. However, the effect of a higher
inclination would be to diminish the ``efficiency'' of the reflection.
This effect would therefore be compensated in the fitting
procedure by an increase of the normalization of the reflection
component, leaving the above conclusions unchanged.
It could also be argued that a different geometry for the reflector,
like that of a (neutral)
broad line region (Yaqoob et al. 1993a, Nandra \& George 1994) or a molecular
torus (Ghisellini, Haardt \& Matt 1994, Krolik et al. 1994) may provide an
alternative description for the observed spectral flatness. However, in all
cases
we would have to face the problem of the absence of a
strong iron line.


Of course if the Fe abundances are reduced the value of the EW can be made to
match the
observed one. However we do not regard this as a realistic possibility as
observational evidence indicates that, at least for some
Seyfert 2 galaxies, the tendency is to have Fe overabundances.
Another alternative possibility comes from
theoretical arguments. Ross \& Fabian (1993) and Zycki \& Czerny (1994)
showed that for a range of ionization stage of iron ions within 15 $\lsimeq
<Fe> \lsimeq 22$, or if iron is fully ionized, the Fe line EW can be very low
even in the presence of strong reflection.
However, it should be emphasized that the presence of highly ionized material
is
compatible only with the case of reflection from an accretion disk
or, possibly, from a broad line region (Netzer 1993), but not from a torus.
Moreover, in the case of ionized reflection, the
substantial lack of photoelectric absorption yields a slope of the
reflection component which is typically the same (or even steeper)
than that of the intrinsic power law.
The Fe K line center energy is also expected at slightly
higher energies than observed if the gas is ionized.
Therefore, the use of ionized reflection instead of neutral reflection
wouldn't be justified in this case.

We also found that a dual-absorber model (section 3.3.3) could explain
the observed flat
spectrum with an intrinsically steep power law ($\Gamma \simeq 1.86 \pm 0.3$).
The deduced spectral shape (Figures 6a or 6b) is very similar to that
expected with a reflection component (Figure 4) but in this case,
a strong emission line is not expected a priori.
The only constrain could come from the absorption edge at the redshifted
energy of 6.94 keV, as expected by neutral absorption. The model predicts
an absorption edge with optical depth $\tau \simeq 0.21^{+0.68}_{-0.11}$ which
is
consistent with the measured upper limit of $\tau \lsimeq 0.16$ (see section
3.4).
Therefore, nonuniform cold absorption provides an adequate explanation for the
observed flat spectrum of NGC5252.
The absorbing material may be interpreted as a distribution of clouds
with nonuniform density, e.g. with higher densities at smaller radii.
Alternatively, it may consist of an absorbing torus with nonuniform
density. In this case, since the absorption is $\sim 10^{22-23}$ cm$^{-2}$,
while predicted columns of absorption tori are $\sim 10^{24-25}$ cm$^{-2}$,
either
(1) the torus is intrinsically optically thin or (2) we are looking through its
rim.

Another possible interpretation for the flat spectrum of NGC5252 is that
it is intrinsically flat.
In a re-analysis of GINGA data, Smith and Done (1995) found several other
Seyfert 2 galaxies with flat power law spectra similar to that observed
in NGC5252, but with slightly higher iron line EW $\sim 100 - 150 $ eV. These
authors also found that at least 4 of the sources are
incompatible with a steep slope, even with the inclusion of a
reflection component.
A recent analysis of combined GINGA and OSSE data by Zdziarski et
al. (1994) indicate that the intrinsic X-ray spectra of some Seyfert 2
galaxies may be substantially harder than those of Seyfert 1s, in
contrast to unified models of AGN. Their slopes are consistent with our
results for NGC5252.

Finally, we would like to comment on the strong resemblance between the
ASCA X-ray spectrum of NGC5252 and the BBXRT and ASCA spectra of NGC4151
(Weaver et al. 1994a, Weaver et al. 1994b).
Both sources reveal a strong (L$_{\rm
X}$(0.1-2.0)$\sim 10^{41}$ erg s$^{-1}$) soft excess below $\sim$ 2
keV, an absorption around $10^{22-23}$ cm$^{-2}$, a flat
continuum ($\Gamma \sim 1.3-1.7$), a high ($\gsimeq 0.90$) covering
fraction and a relatively weak iron line ($\sim 50-150$ eV) (Yaqoob et
al. 1993b, Weaver et al. 1994b).  Both objects are relatively low luminosity
Seyfert galaxies with L$_{\rm X}$(2-10 keV)$ \sim 2 \times 10^{43}$ erg
s$^{-1}$ (NGC5252) and L$_{\rm X}$(2-10 keV)$ \sim 10^{43}$ erg
s$^{-1}$ (NGC4151; Seyfert 1.5).
These strong similarities suggest that the hard spectrum of
NGC5252 may be intrinsically flat, as seems to be the case for NGC4151.

A better understanding of these Seyfert galaxies
with flat spectra is of fundamental importance, not only because
the problem has direct implications for unified models, but also
because they could help explaining the so-called ``spectral paradox''
(Boldt 1987), i.e. the difference between the X-ray spectra of AGN and
that of the CXB.




\pn

 \section{Conclusion}

The analysis of the X-ray spectrum of NGC5252 shows that a single
absorbed power law alone can describe the 0.7-10 keV GIS spectrum but
there is evidence of a strong soft excess in the 0.4-10
keV SIS spectrum.  We argue against a thermal origin of the soft excess
because there is no indication of starburst activity in NGC5252 and we do
not observe the expected soft emission lines produced by a plasma
at T $\sim 10^{6}$ K.

The best description for the 0.4-10 keV spectrum is provided by a partial
covering model or a scattering model, both models being equally
compatible with the current data.
The main results of the spectral analysis are a very flat ($\Gamma
\simeq 1.45 \pm 0.23$) power law, a high covering fraction ($C_f \simeq
0.96 \pm 0.01$) from thick ($N_{\rm H} \simeq 4.3
\pm 0.6 \times 10^{22}$ cm$^{-2}$) material and a weak (EW$\simeq
90 \pm 60$ eV) iron line. The scattering model gives the same best-fit
parameters, with a fraction $\sim 4$\% of scattered light.

The flat spectral shape is incompatible with current theories of
Seyfert unification schemes which predict steeper intrinsic power
laws of $\Gamma \sim 1.8$ - 2.1.
The flat slope {\bf and} weak Fe K iron line measured in NGC5252
make the reflection hypothesis unlikely, but it cannot be ruled out.
The inclusion of a warm absorber, instead of a neutral absorber
to explain the flat spectrum gives systematic
deviations at higher energies yielding a fit
statistically worse than with only partial covering.
A complex partial covering model with two
different column densities (composite dual-absorber model) with a
canonical steep ($\Gamma \sim 1.9$) power law offers a potential
explanation for the flatness of the spectrum.
An alternative interpretation of the present data is that the
measured X-ray spectrum of NGC5252 is intrinsically flat.

 \section{Acknowledgements}

 \ms The authors wish to thank the referee, Kim Weaver, for helpful and
constructive comments, all the ASCA team members for making
this observation possible and K. Leighly for a carefull reading of the
manuscript. M.C. thanks A. Comastri, S. Molendi, Y. Ogasaka, S.
Ueno and T. Yamada for helpful discussions. M.C. also acknowledges
financial support from the Science and Technology Agency of Japan (STA
fellowship), hospitality from the RIKEN Institute and support from the
European Union.  G.G.C.P. acknowledges partial financial support from
MURST and ASI.
\ms

\vfill\eject

\section{References}

\pn
Antonucci, R.R.J., 1993, ARA\&A, 31, 473

\pn
Antonucci, R.R.J. \& Miller, J.S., 1985, ApJ, 297, 621


\pn
Arnaud, K.A. et al., 1991, XSPEC User's Guide, ESA TM-09

\pn
Avni, Y., 1976, ApJ, 210, 642

\pn
Awaki, H. et al., 1991, PASJ, 43, 195

\pn
Boldt, E. 1987, Phys. Repts., 146, 215

\pn
David, L.P., Jones, C. \& Forman, W., 1992, ApJ, 388, 82

\pn
Dickey, J.M. \& Lockman, F.J., 1990, ARA\&A, 28, 215

\pn
Fabian, A.C. et al., 1994, PASJ, 46, L59

\pn
Ferland, G.J., 1991, OSU Internal report, N. 91-01

\pn
Fukazawa, Y. et al., 1994, PASJ, 46, L141

\pn
George, I.M. \& Fabian, A.C., 1991, MNRAS, 249, 352

\pn
Ghisellini, G., et al., 1991, MNRAS, 248, 14

\pn
Ghisellini, G., Haardt, F. \& Matt, G., 1994, MNRAS, 267, 743

\pn
Guilbert, P.W., \& Rees, M.J., 1988, MNRAS, 233, 475

\pn
Holt, S.S. et al., 1980, ApJ, 241, L13



\pn
Kay, L., 1994, ApJ, 430, 196

\pn
Krolik, J.H., Madau, P. \& Zycki, P.T., 1994, ApJ, 420, L57

\pn
Kotilanien, J. \& Prieto, M.A., 1994, A\&A, 295, 649

\pn
Koyama, K. et al., 1989, PASJ, 41, 731

\pn
Krolik, J.H. \& Kriss, G.A., 1995, preprint.

\pn
Kunieda, H. et al. 1994, in ``New Horizon of X-ray Astronomy'', ed. F. Makino
\& T. Ohashi, (Tokyo: Universal Academy Press), p317

\pn
Lampton, M., Margon, B. \& Bowyer, S., 1976, ApJ, 208, 177

\pn
Leahy, D.A. \& Creighton, J., 1993, MNRAS, 263, 314

\pn
Lightman, A.P. \& White, T.R., 1988, ApJ, 335, 57

\pn
Makishima, K. 1986, Lecture Notes in Physics, 266, 249

\pn
Marshall, F.E. et al., 1993, ApJ, 405, 168

\pn
Matsuoka, M. 1994, in ```New Horizon of X-ray Astronomy'', ed. F. Makino
\& T. Ohashi, (Tokyo: Universal Academy Press), p305

\pn
Miller, J.S. \& Goodrich, R.W., 1990, ApJ, 355, 456

\pn
Morrison, R. \& McCammon, D., 1983, ApJ, 270, 119

\pn
Nandra, K. \& George, I.M., 1994, MNRAS, 267, 974

\pn
Nandra, K. \& Pounds, K.A., 1994, MNRAS, 268, 405

\pn
Netzer, H., 1993, ApJ, 411, 594

\pn
Otani, C. \& Dotani, T., 1994, Asca News n.2, 25

\pn
Prieto, M.A. \& Freudling, W., 1993, ApJ, 418, 668

\pn
Prieto, M.A. \& Freudling, W., 1995, MNRAS, submitted

\pn
Raymond, J.C. \& Smith, B.W., 1977, ApJS, 35, 419

\pn
Reynolds, C.S. et al., 1994, MNRAS, 268, L55

\pn
Ross, R.R. \& Fabian, A.C., 1993, MNRAS, 261, 74

\pn
Ruiz,M., Rieke, G. \& Schmidt, G., 1994, ApJ, 423, 608

\pn
Smith, D.A., \& Done, C., 1995, submitted to MNRAS

\pn
Tadhunter, C. \& Tsvetanov, Z., 1989, Nature, 341, 422

\pn
Tran, H.D., Miller, J.S. \& Kay, L.E., 1992, ApJ, 397, 452


\pn
Ueno, S. et al., 1994, PASJ, 46, L71

\pn
Unger, S.W. et al, 1987, MNRAS, 228, 671

\pn
Weaver, K.A., Arnaud, K.A. \& Mushotzky, R.F., 1995, ApJ, in press

\pn
Weaver, K.A. et al., 1994a, ApJ, 423, 621

\pn
Weaver, K.A. et al., 1994b, ApJ, 436, L27

\pn
Wilson, A.S. \& Tsvetanov, Z.I., 1994, A.J., 107, 1227

\pn
Yaqoob, T. et al., 1993a, ApJ, 416, L5

\pn
Yaqoob, T. et al., 1993b, MNRAS, 262, L435

\pn
Zdziarski, A.A. et al., 1994, ApJL, 438, L63

\pn
Zycki, P.T. \& Czerny, B., 1994, MNRAS, 266, 653

\vfill\eject

\vskip 1truecm
\centerline{\bf Table 1 : Selection criteria}
\begin{center}
\begin{tabular}{cc}
\hline
\hline
\multicolumn{1}{c}{GIS} &
\multicolumn{1}{c}{SIS} \\
\hline
SAA=0 & SAA=0 \\
\hline
COR\_MIN$>$ 7 & COR\_MIN$>$ 6 \\
\hline
ELV\_MIN$>$ 5 & ELV\_MIN$>$ 5 \\
\hline
-- & BR\_EARTH$>$ 25 \\
\hline
-- & T\_DY\_NT$>$ 100 \\
\hline
\hline
\end{tabular}
\end{center}
\pn
SAA: South Atlantic Anomaly
\pn
COR\_MIN: Cut-off rigidity (GeV/c)
\pn
ELV\_MIN: Minimum elevation angle from earth (degree)
\pn
BR\_EARTH: Minimum bright earth angle (degree)
\pn
T\_DY\_NT: Time after a day/night transition (s)

\vfill\eject

\vskip 1truecm
\centerline{\bf Table 2 : Exposure times and count rates}
\begin{center}
\begin{tabular}{ccccc}
\hline
\hline
\multicolumn{1}{c}{} &
\multicolumn{1}{c}{exposure} &
\multicolumn{1}{c}{Source + Backg.} &
\multicolumn{1}{c}{Backg.} &
\multicolumn{1}{c}{Source}\\
\multicolumn{1}{c}{} &
\multicolumn{1}{c}{(s)} &
\multicolumn{1}{c}{(counts)} &
\multicolumn{1}{c}{(counts/s)} &
\multicolumn{1}{c}{(counts/s)}\\
\hline
GIS2 & 70970 & 4011 & 0.011 & 0.045 \\
\hline
GIS3 & 70747 & 5194 & 0.012 & 0.061 \\
\hline
SIS0 & 65154 & 4235 & 0.010 & 0.055 \\
\hline
SIS1 & 64448 & 3736 & 0.008 & 0.050 \\
\hline
\hline
\end{tabular}
\end{center}

\vfill\eject

\vskip 1truecm
\centerline{\bf Table 3: GIS and SIS - Single Power Law}
\begin{center}
\begin{tabular}{ccccc}
\hline
\hline
\multicolumn{1}{c}{} &
\multicolumn{1}{c}{$N_{\rm H}$} &
\multicolumn{1}{c}{$\Gamma$} &
\multicolumn{1}{c}{F$^{a}_{\rm X}$(0.7-10 keV)} &
\multicolumn{1}{c}{$\chi^{2}_{red}/$d.o.f.} \\
\multicolumn{1}{c}{} &
\multicolumn{1}{c}{($10^{22}$ cm$^{-2}$)} &
\multicolumn{1}{c}{} &
\multicolumn{1}{c}{($10^{-12}$ erg cm$^{-2}$ s$^{-1}$)} &
\multicolumn{1}{c}{} \\
\hline
GIS2&${3.76^{+0.84}_{-0.75}}$&${1.35 ^{+0.25}_{-0.24}}$ &8.23 &0.83/118 \\
GIS3&${3.91^{+0.71}_{-0.63}}$&${1.43^{+0.20}_{-0.20}}$ &9.77 &0.99/146 \\
BOTH& ${3.83^{+0.54}_{-0.49}}$ & ${1.39^{+0.16}_{-0.15}}$ & 9.00 &1.03/267 \\
\hline
\hline
SIS0& ${2.13^{+0.73}_{-0.52}}$&${0.84^{+0.31}_{-0.24}}$ &9.00 &1.39/165 \\
SIS1& ${2.52^{+0.73}_{-0.52}}$&${1.05^{+0.32}_{-0.30}}$ &9.46&1.36/145 \\
BOTH&${2.34^{+0.48}_{-0.44}}$  & ${0.95^{+0.20}_{-0.20}}$ & 9.20 &1.38/313 \\
\hline
\hline
\end{tabular}
\end{center}
\pn
$\ ^{a}$ corrected for absorption.
\pn
Note: Intervals are at 90 \% confidence for 2 interesting parameters.

\vfill\eject.

\vskip 1truecm
\centerline{\bf Table 4: SIS -- Two component models}
\begin{center}
\begin{tabular}{ccccccc}
\hline
\hline
\multicolumn{1}{c}{} &
\multicolumn{1}{c}{Ab.} &
\multicolumn{1}{c}{$kT$} &
\multicolumn{1}{c}{$N_{\rm H}$} &
\multicolumn{1}{c}{$\Gamma$} &
\multicolumn{1}{c}{L$^{a}_{\rm X}$(0.1-2.0 keV)} &
\multicolumn{1}{c}{$\chi^{2}_{red}$/d.o.f.} \\
\multicolumn{1}{c}{} &
\multicolumn{1}{c}{} &
\multicolumn{1}{c}{(keV)} &
\multicolumn{1}{c}{($10^{22}$ cm$^{-2}$)} &
\multicolumn{1}{c}{} &
\multicolumn{1}{c}{($10^{41}$ erg s$^{-1}$)} &
\multicolumn{1}{c}{} \\
\hline
black body + abs(PL)&...& ${0.18^{+0.05}_{-0.03}}$&${3.04^{+0.74}_{-0.53}}$
&${1.17 ^{+0.20}_{-0.16}}$ & 3.86 &0.94/311 \\
bremss. + abs(PL)&...& ${0.80^{+1.23}_{-0.33}}$&${3.45^{+0.83}_{-0.62}}$
&${1.26 ^{+0.26}_{-0.24}}$ & 6.47 &0.93/311 \\
R-S$^{b}$ + abs(PL)&$<0.01$& ${0.77^{+1.18}_{-0.24}}$&${3.46^{+0.84}_{-0.66}}$
&${1.26^{+0.23}_{-0.18}}$ & 6.44 &0.92/310 \\
\hline
\hline
\end{tabular}
\end{center}
\pn
$\ ^{a}$ calculated with only the soft component.
\pn
$\ ^{b}$ R-S = Raymond-Smith (with abundances Ab).
\pn
Note: Intervals are at 90 \% confidence for 3 interesting parameters.

\vfill\eject

\vskip 1truecm
\centerline{\bf Table 5a: SIS -- Partial covering model}
\begin{center}
\begin{tabular}{ccccc}
\hline
\hline
\multicolumn{1}{c}{$N_{\rm H}$} &
\multicolumn{1}{c}{$C_{F}$} &
\multicolumn{1}{c}{$\Gamma$}&
\multicolumn{1}{c}{F$^{a}_{\rm X}$(0.4-10 keV)} &
\multicolumn{1}{c}{$\chi^{2}_{red}/d.o.f.$}\\
\multicolumn{1}{c}{($10^{22}$ cm$^{-2}$)} &
\multicolumn{1}{c}{} &
\multicolumn{1}{c}{} &
\multicolumn{1}{c}{($10^{-12}$ erg cm$^{-2}$ s$^{-1}$)} &
\multicolumn{1}{c}{} \\
\hline
4.33$^{+0.66}_{-0.61}$ & 0.96$^{+0.01}_{-0.02}$ & 1.45$^{+0.23}_{-0.23}$ &
11.69 & 0.95/312\\
\hline
\hline
\end{tabular}
\end{center}
\centerline{\bf Table 5b: SIS -- Scattering model}
\begin{center}
\begin{tabular}{ccccc}
\hline
\hline
\multicolumn{1}{c}{$N_{\rm H}$} &
\multicolumn{1}{c}{A(1)/A(2)$^{b}$} &
\multicolumn{1}{c}{$\Gamma$}&
\multicolumn{1}{c}{F$^{a}_{\rm x}$(0.4-10 keV)} &
\multicolumn{1}{c}{$\chi^{2}_{red}/d.o.f.$} \\
\multicolumn{1}{c}{($10^{22}$ cm$^{-2}$)} &
\multicolumn{1}{c}{} &
\multicolumn{1}{c}{} &
\multicolumn{1}{c}{($10^{-12}$ erg cm$^{-2}$ s$^{-1}$)} &
\multicolumn{1}{c}{} \\
\hline
4.39$^{+0.68}_{-0.64}$ & 0.046 & 1.46$^{+0.23}_{-0.23}$ & 11.69 & 0.95/312\\
\hline
\hline
\end{tabular}
\end{center}
\pn
$\ ^{a}$ corrected for absorption.
\pn
$\ ^{b}$ A(1)/A(2) = ${\rm normalization\ of\ the\ scattered\
conponent \ at\ 1\ keV \over{\rm normalization \ of\ the\ continuum\ at\ 1
\ keV}}$
\pn
Note: Intervals are at 90 \% confidence for 3 interesting parameters.

\vfill\eject

\normalsize
\vskip 1truecm
\centerline{\bf Table 6: SIS -- Partial covering + reflection}
\begin{center}
\begin{tabular}{cccccccc}
\hline
\hline
\multicolumn{1}{c}{$N_{\rm H}$} &
\multicolumn{1}{c}{$C_{F}$} &
\multicolumn{1}{c}{$A_{\rm pl}^a$} &
\multicolumn{1}{c}{$A_{\rm refl}^a$} &
\multicolumn{1}{c}{$\Gamma_{\rm int}$} &
\multicolumn{1}{c}{$\chi^{2}_{red}/d.o.f.$} \\
\multicolumn{1}{c}{($10^{22}$ cm$^{-2}$)} &
\multicolumn{1}{c}{} &
\multicolumn{1}{c}{} &
\multicolumn{1}{c}{} &
\multicolumn{1}{c}{} &
\multicolumn{1}{c}{} \\
\hline
4.61$^{+0.40}_{-0.39}$&0.97$^{+0.01}_{-0.01}$&2.19&15.6$^{+16.3}_{-12.3}$&1.91$^{+0.33}_{-0.33}$&0.92/311\\
\hline
\hline
\end{tabular}
\end{center}
\pn
$\ ^{a}$ normalization at 1 keV in units of $10^{-3}$ photons cm$^{-2}$
s$^{-1}$ keV$^{-1}$.
\pn
Note: Intervals are at 90 \% confidence for 1 interesting parameter.

\vfill\eject

\normalsize
\vskip 1truecm
\centerline{\bf Table 7: SIS -- Composite dual-absorber model}
\begin{center}
\begin{tabular}{ccccccc}
\hline
\hline
\multicolumn{1}{c}{$N_{\rm H1}$} &
\multicolumn{1}{c}{A$^{a}$(1)} &
\multicolumn{1}{c}{$C_{F}$} &
\multicolumn{1}{c}{$\Gamma$} &
\multicolumn{1}{c}{$N_{\rm H2}$} &
\multicolumn{1}{c}{A$^{a}$(2)} &
\multicolumn{1}{c}{$\chi^{2}_{red}/d.o.f.$} \\
\multicolumn{1}{c}{($10^{22}$ cm$^{-2}$)} &
\multicolumn{1}{c}{} &
\multicolumn{1}{c}{} &
\multicolumn{1}{c}{} &
\multicolumn{1}{c}{($10^{22}$ cm$^{-2}$)} &
\multicolumn{1}{c}{} &
\multicolumn{1}{c}{} \\
\hline
4.82$^{+0.40}_{-0.37}$& 2.17&
0.98$^{+0.01}_{-0.01}$&1.86$^{+0.26}_{-0.31}$&62$^{+64}_{-25}$ & 1.75 &
0.92/311\\
\hline
\hline
\end{tabular}
\end{center}
\pn
$\ ^{a}$ normalization at 1 keV in units of $10^{-3}$ photons cm$^{-2}$
s$^{-1}$ keV$^{-1}$.
\pn
Note: Intervals are at 90 \% confidence for 1 interesting parameter.

\vfill\eject

\normalsize
\vskip 1truecm
\centerline{\bf Table 8: SIS -- Fe K line best-fit parameters}
\begin{center}
\begin{tabular}{ccccccc}
\hline
\hline
\multicolumn{1}{c}{Continuum} &
\multicolumn{1}{c}{E} &
\multicolumn{1}{c}{$\sigma$} &
\multicolumn{1}{c}{EW} \\
\multicolumn{1}{c}{model} &
\multicolumn{1}{c}{(keV)} &
\multicolumn{1}{c}{(eV)} &
\multicolumn{1}{c}{(eV)} \\
\hline
Raymond-Smith + Power law & $6.25^{+0.05}_{-0.16}$& 0 (fixed)& $\lsimeq 120$\\
\hline
Partial covering & $6.25^{+0.05}_{-0.07}$& 0 (fixed)& $90^{+63}_{-49}$\\
& $6.25^{+0.05}_{-0.03}$ & $\lsimeq 118$& $100^{+62}_{-59}$\\
\hline
Partial covering + Reflection & $6.24^{+0.07}_{-0.08}$ & 0 (fixed) &
$78^{+44}_{-53}$\\
\hline
\hline
\end{tabular}
\end{center}
\pn
Note: Intervals are at 90 \% confidence for 1 interesting parameter.

\vfill\eject

\centerline{\bf Figure captions}
\par\noindent
Figure 1: SIS spectrum and associated residuals obtained from
the simultaneous joint fit of SIS0 and
SIS1 instrument data with a single absorbed power law.
The solid line represents the best-fit power law modelled to the spectrum
between 0.4 - 10 keV.
A soft excess below 1.2 keV is clearly evident.
\bs
Figure 2:
Unfolded SIS spectrum with the partial covering model. The solid line
represents the best-fit partial covering model.
\bs
Figure 3:
Confidence contours for photon index vs column density ({\it top}) and
covering fraction vs column density ({\it bottom}) for the best fit obtained
with the partial covering model. Contours are at 68\%, 90\% and 99\% confidence
levels.
\bs
Figure 4:
SIS best-fit model and associated residuals with
the partial covering + reflection component model.
\bs
Figure 5: SIS best-fit model and associated residuals with the scattering +
warm absorber model.
\bs
Figure 6: (a) SIS best-fit model and associated residuals with
the partial covering + ``dual-absorber'' model. (b) SIS best-fit model and
associated residuals with the scattering + ``dual-absorber'' model.

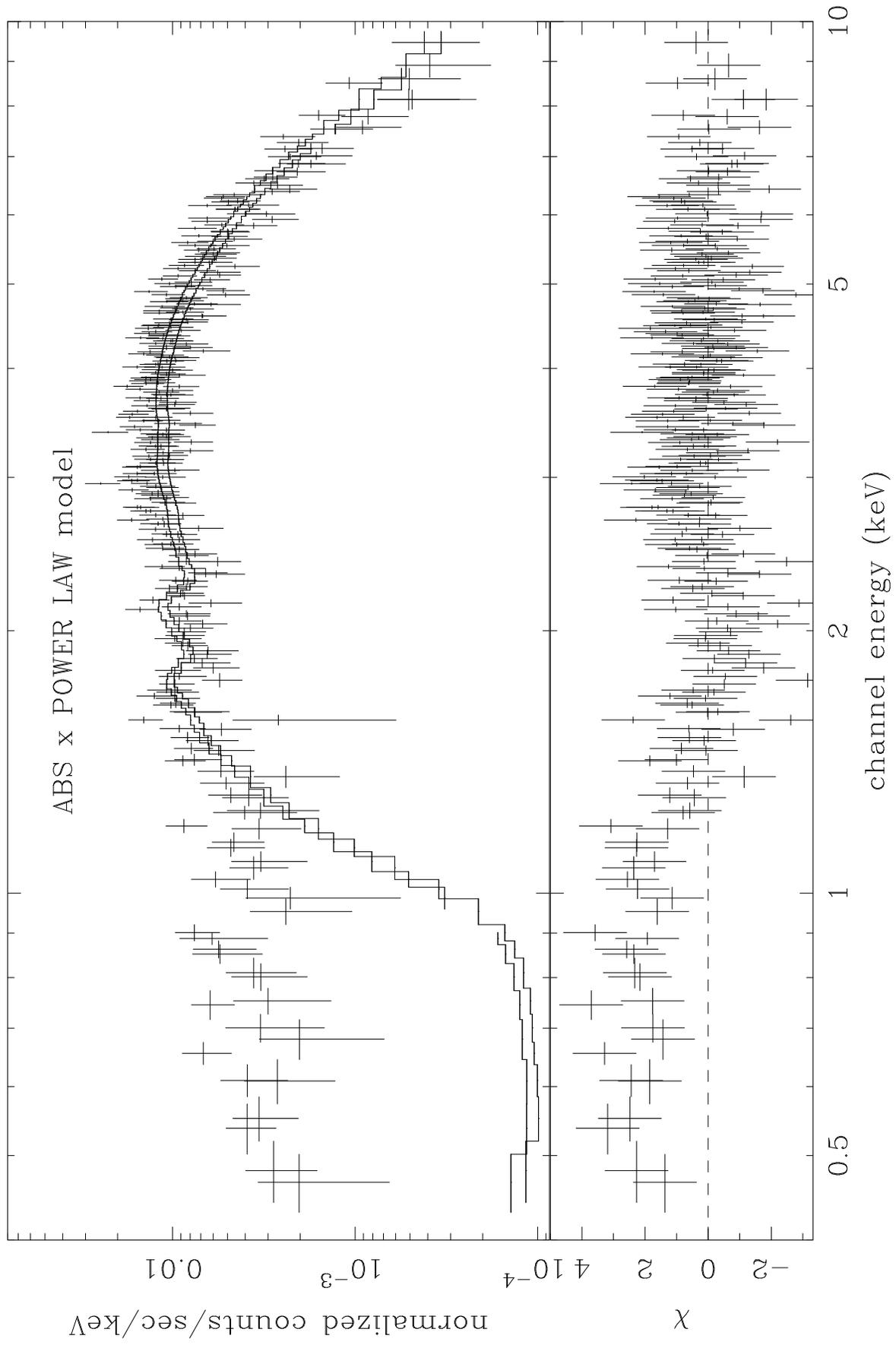
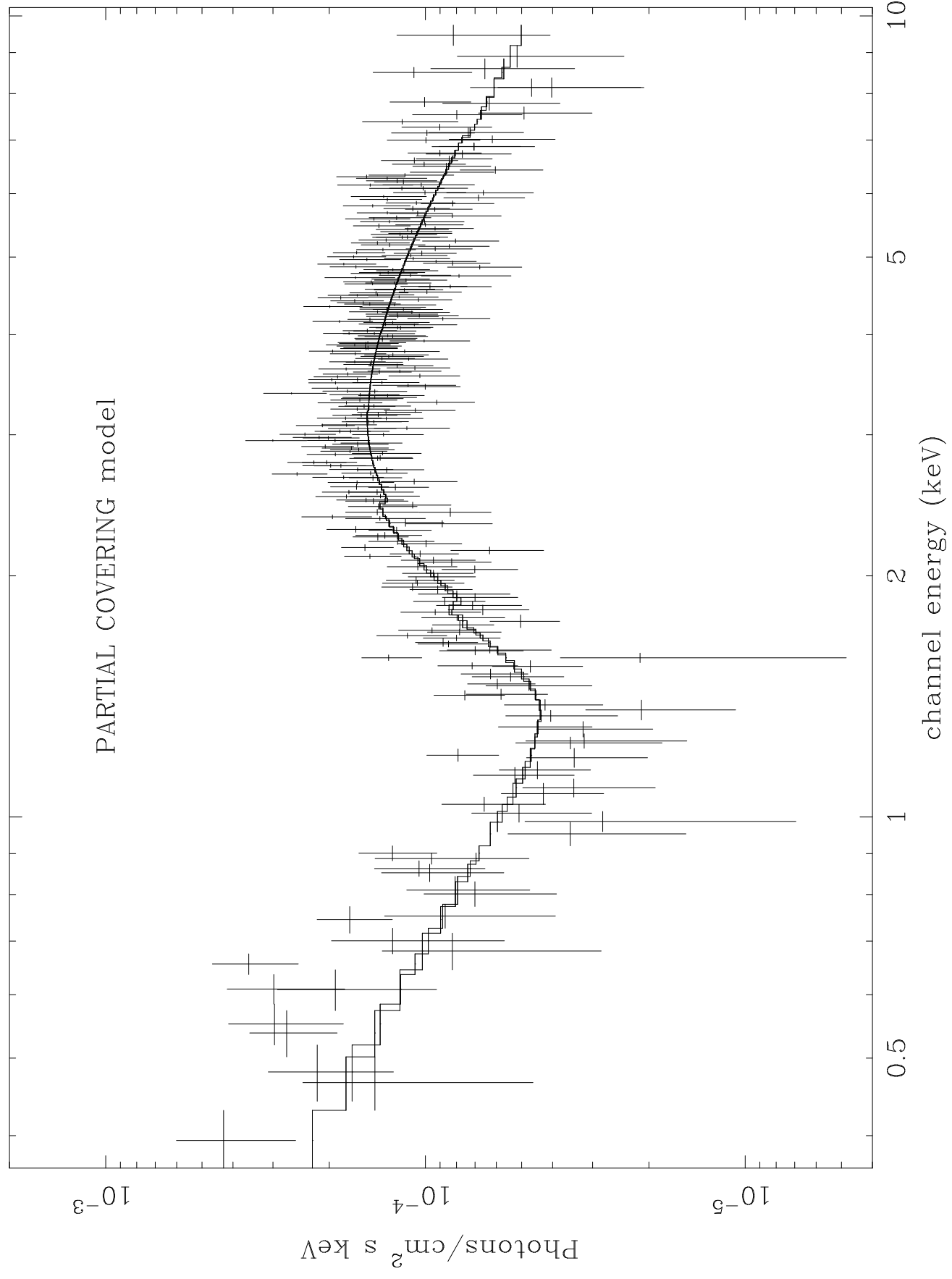
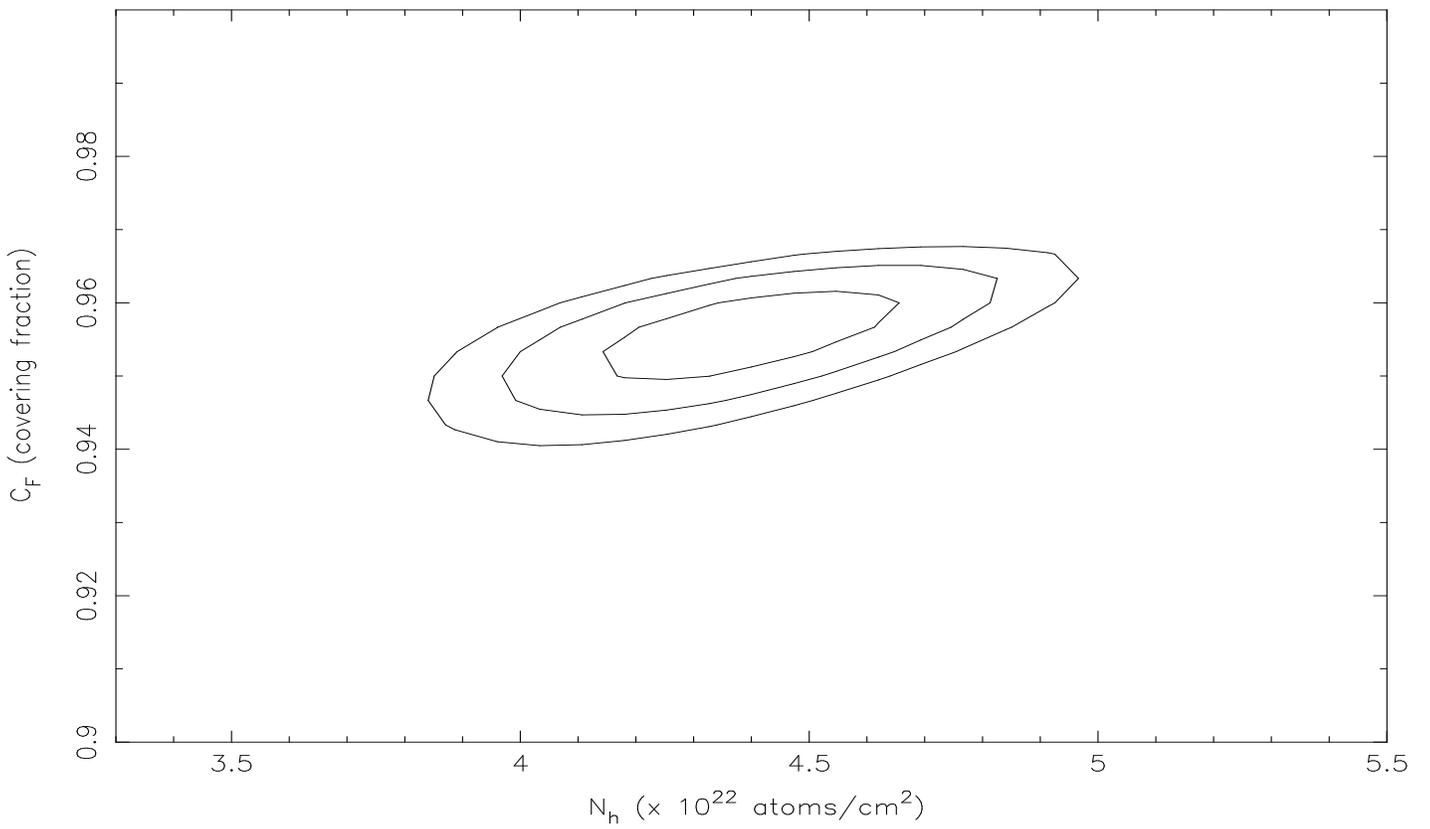
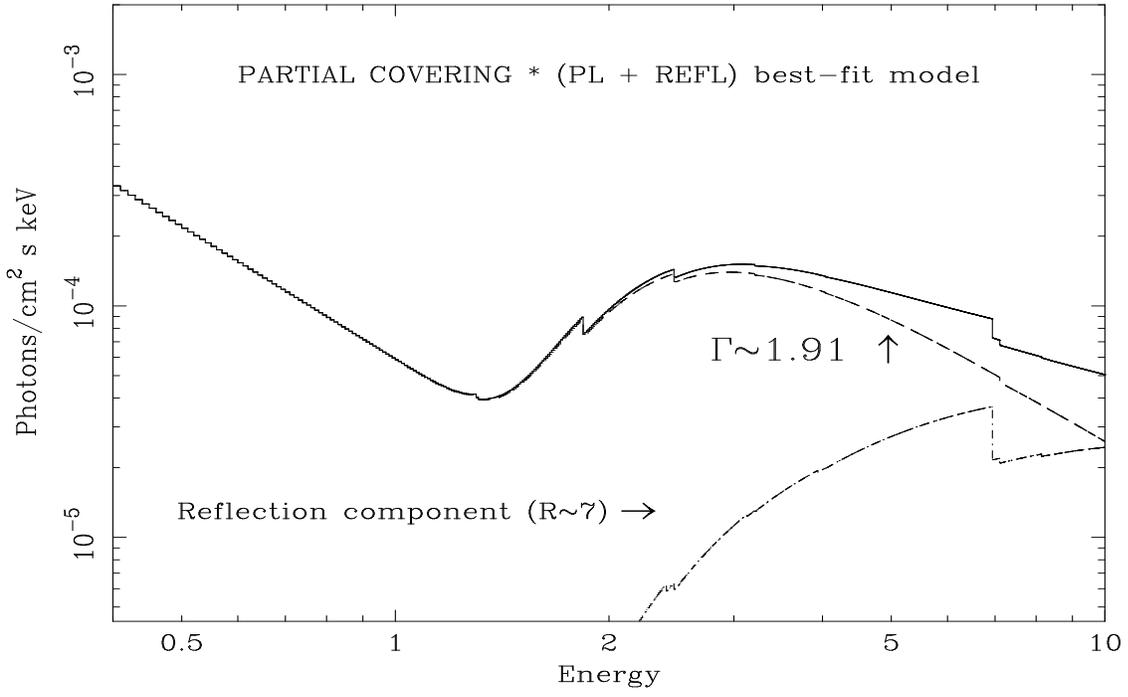
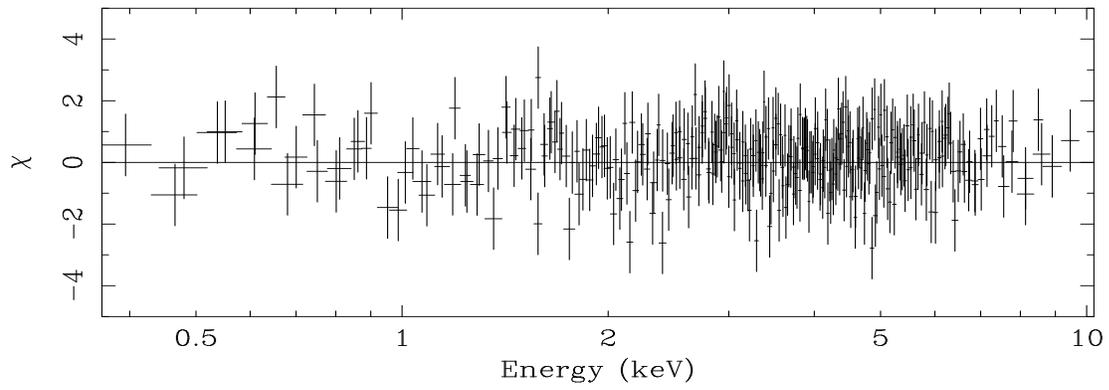
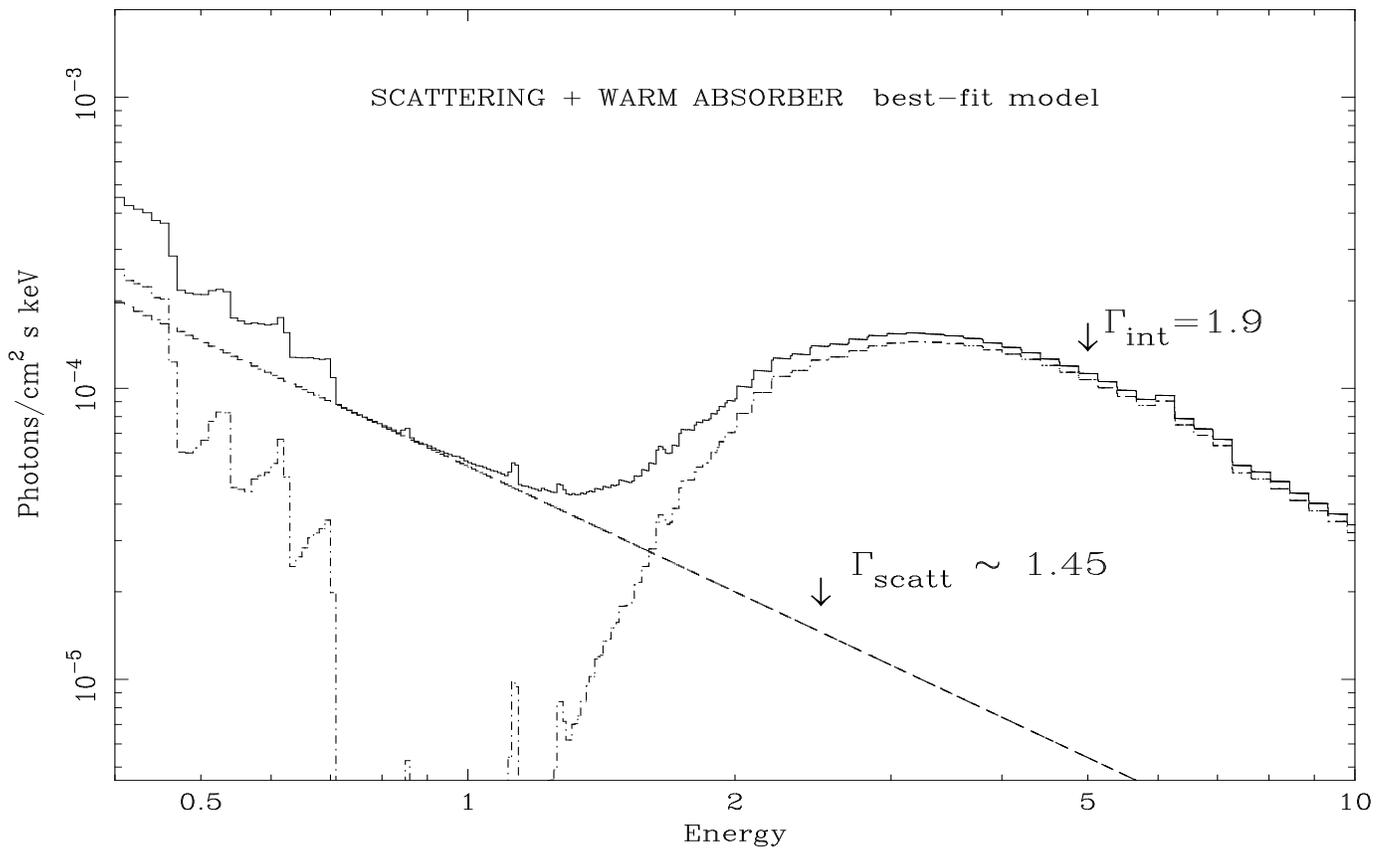
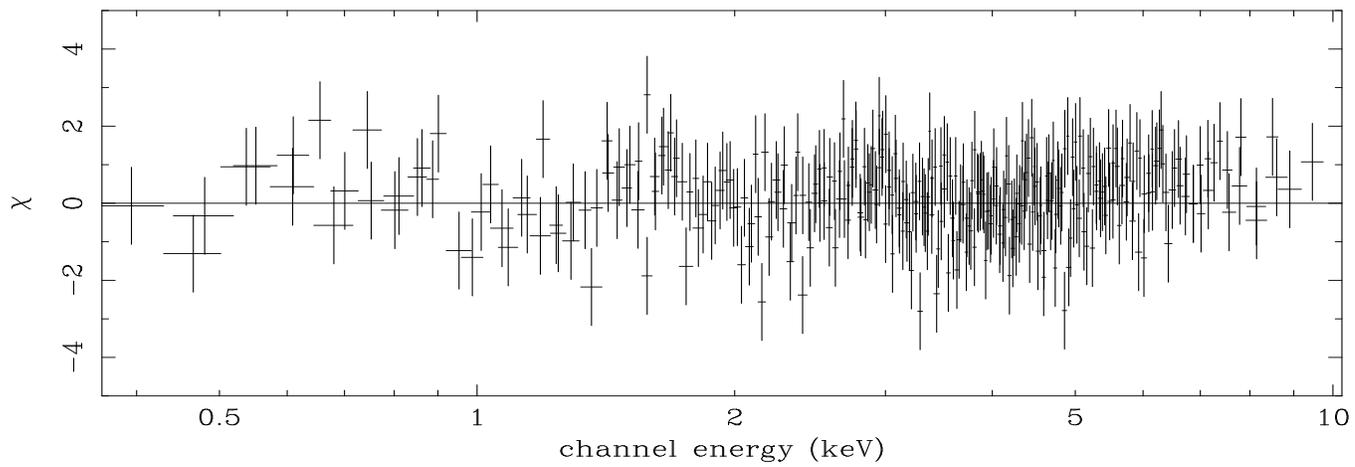
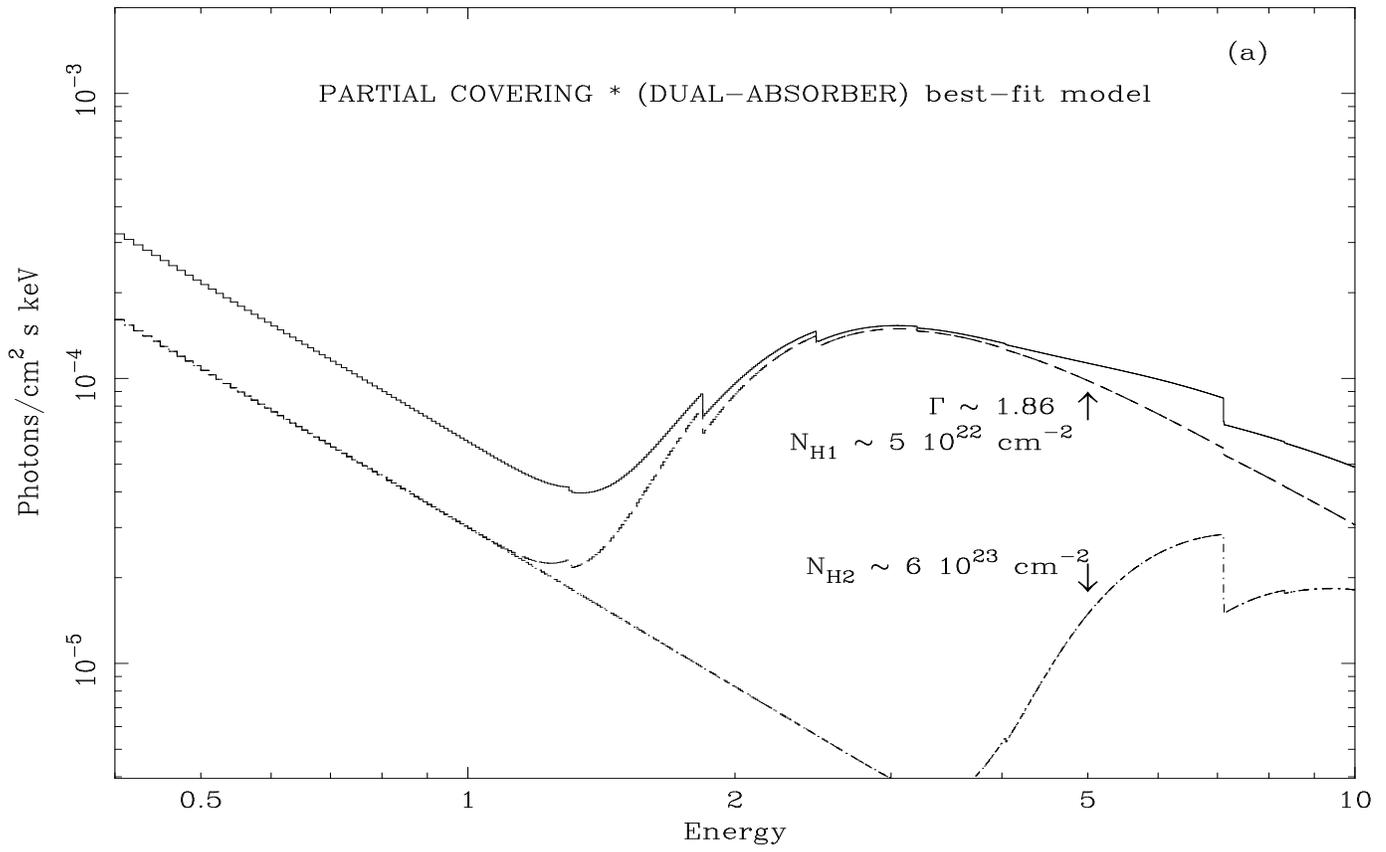
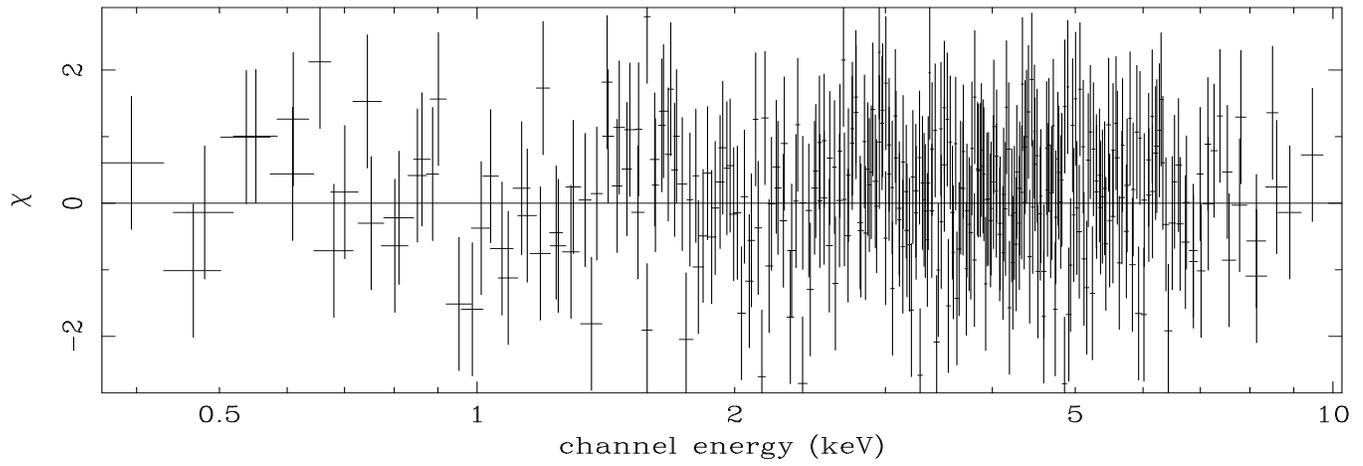
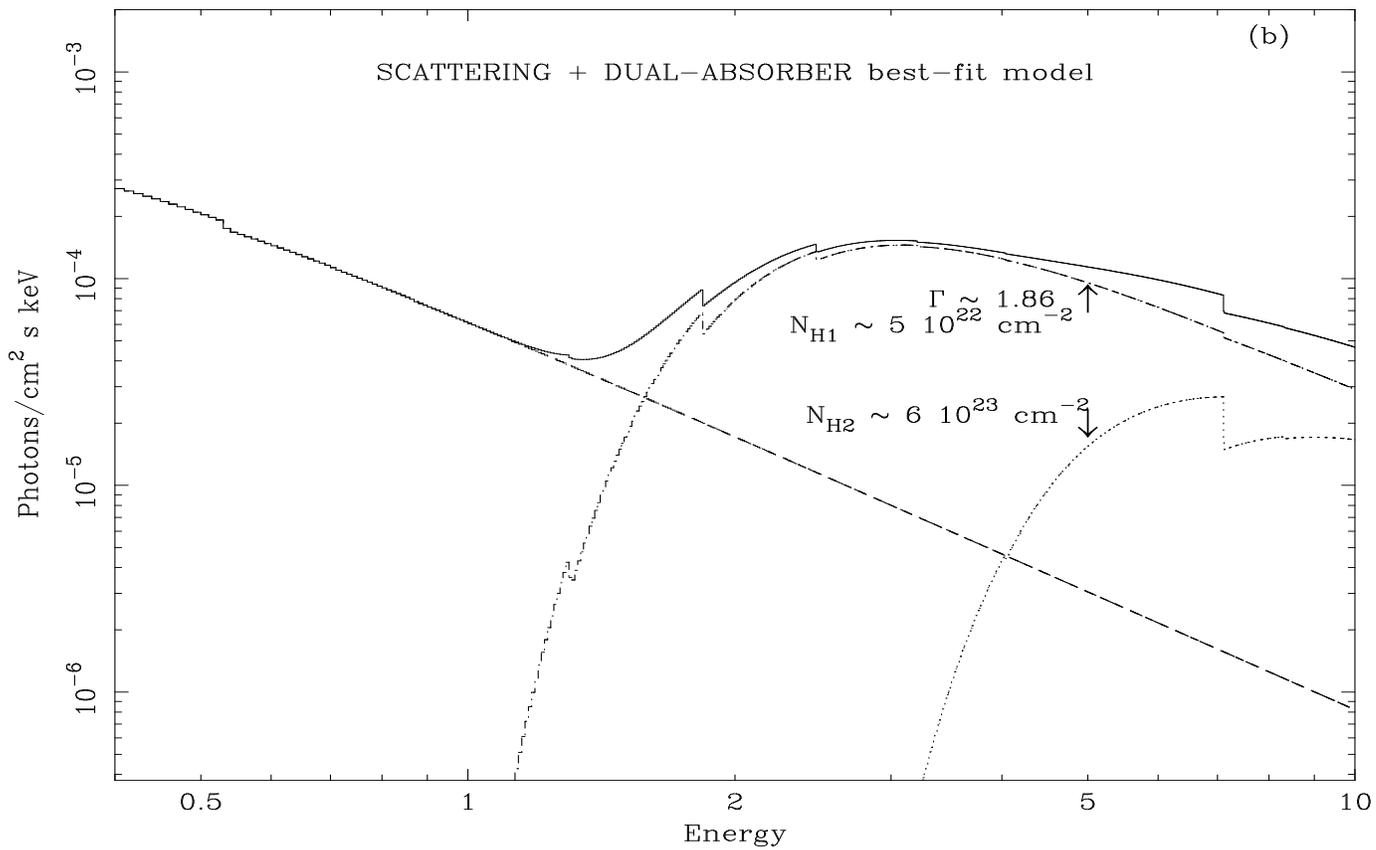
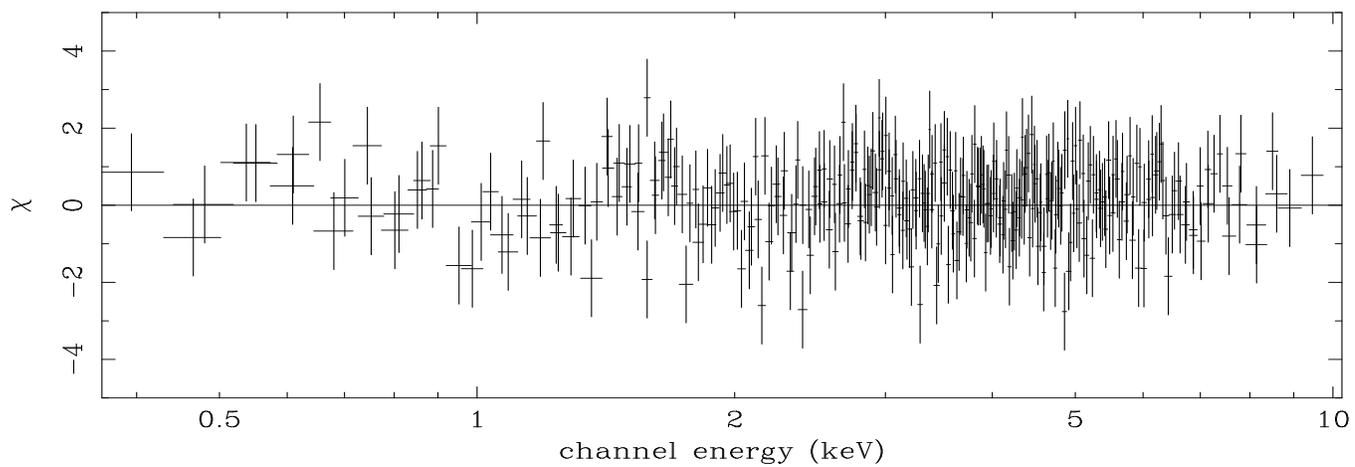
\end{document}